# Social Live-Streaming Use & Well-being:

# Examining Participation, Financial Commitment, Social Capital, and Psychological Well-being on Twitch.tv

Grace H. Wolff[1] and Cuihua Shen[2]

Department of Communication, University of California Davis

Davis, California, United States of America

[1]ghwolff@ucdavis.edu, ORCID 0000-0001-9337-9117

[2]cuishen@ucdavis.edu, ORCID 0000-0003-1645-8211

**Abstract**

This study examines how active participation, financial commitment, and passive participation in the leading social live-streaming service, Twitch.tv, relate to individuals' psychological well-being. The three dimensions of social capital—structural, relational, and cognitive—as well as parasocial relationship are explored as mediators. Cross-sectional survey data from 396 respondents was analyzed by comparing two fully saturated structural equation models. Findings indicate actively participating in a favorite streamers' Chat is positively associated with increased well-being. Structural social capital, or having more social interaction ties, positively mediates the relationship between active participation and well-being, as well as financial commitment and well-being. Greater cognitive social capital, or shared values and goals with a favorite streamer, is related to decreased well-being. Parasocial relationship does not significantly mediate the relationship between use and well-being. Our results demonstrate the importance of tangible social ties over the perceived relationships or identification with a favorite streamer.

**Social Live-Streaming Use & Well-being:**

**Examining Participation, Financial Commitment, Social Capital, and Psychological**

**Well-being on Twitch.tv**

Social live-streaming services (SLSSs) are social media entertainment platforms that combine elements of connectivity and community typical in social media, with entertainment ecosystems of producers and consumers (Johnson & Woodcock, 2019a). With the global streaming market increasing at 28.1% annually and projections to reach almost $250,000 million by 2027 (Market Research Future, 2021), it is no surprise that young adults now spend more time viewing live-streaming content than traditional cable shows (Hu, Zhang, & Wang, 2017). SLSSs enable users to broadcast or "live-stream" their own content in real time to interactive audiences who can in turn influence the broadcast with their suggestions, comments, and financial contributions (Carter & Egliston, 2018; Scheibe et al., 2016). With a focus on amateur content production and synchronous communication that facilitate greater intimacy and sociability, SLSSs support microcelebrity sub-communities and cultures that enable streamers to monetize their content and receive direct financial contributions from viewers (Johnson & Woodcock, 2019b). Considering how SLSSs have reshaped interactive experiences, social engagement, and sense of belonging in online communities (Wohn, Freeman, & McLaughlin, 2018), the question remains as to how SLSS use relates to individuals' psychological well-being (PWB).

While the past decade of research has primarily examined social media use (SMU) and PWB in platforms such as Facebook (see Meier & Reinecke, 2020 for review), there is early research exploring the relationship between SMU and PWB in SLSSs. However, these studies have primarily been grounded in e-Sports or video game live-streaming contexts (Kim & Kim, 2020; Chen & Chang, 2019) or have focused on problematic or addictive use and their relationship to depression and loneliness (Chen & Chang, 2019; Wan & Wu, 2020), while failing to consider a wide variety of SLSS genres beyond gaming and the broad spectrum of users. Additionally, these SLSS studies do not highlight the explanatory role of social capital, which has long-been established as a key mechanism influential to psychological well-being (Ellison, Steinfeld, & Lampe, 2007; Joseph, 2020; Kim & Shen, 2020). To address these gaps, this paper extends existing research on the relationship between media use and PWB by sampling viewers across a variety of streaming genres, measuring their positive psychological well-being, and being the first to examine the role of social capital in SLSS contexts.

As SLSS viewers' use on these platforms are motivated by their perceived relationship with a streamer (Hu, Zhang, & Wang, 2017; Lim, Choe, Zhang, & Noh, 2020), we examine viewers' social capital and parasocial relationships to understand how viewers' usage on the leading SLSS, Twitch.tv, relate to their psychological well-being. Findings from 396 survey respondents indicate that SLSS users derive well-being benefits from actively participating in the Chat of their favorite streamers' broadcasts, and that having close social ties on Twitch may support the well-being of actively participatory and financially committed members. While SLSS environments provide optimal conditions for parasocial relationships to emerge, we found that these asymmetrical relationships, or the perceived relational closeness with a streamer, are not substantially related to viewers' well-being. Prior studies have examined social capital in light of network structure and related outcomes (Williams, 2006), however as our study demonstrates, measuring social capital as a social resource may be more suitable for consumer-generated content communities (Bründl & Hess, 2016; Jeong, Ha, & Lee, 2020).

**Social Interactions in SLSSs**

Below we outline SLSSs' unique network structure and interaction affordances that set it apart from other social media platforms. A social network is a set of people or entities connected by some kind of social relationship, such as a friendship, colleague, or information-provider (Garton, Haythornthwaite, & Wellman, 1997). These relationships are described as "ties," which vary in direction and strength. A tie can be symmetric (ie: Person A and Person B are mutually connected) or asymmetric or uni-directional (ie: Person A "follows" Person B only) and can also vary in strength, or the amount of time, emotional intensity and intimacy of a relationship, often delineated as "strong" and "weak" ties (Granovetter, 1973). Weak ties are characterized by relatively infrequent interactions and low levels of emotional closeness (Putnam, 2001) which help facilitate information diffusion and access to diverse resources, helping to integrate smaller network clusters into larger social networks (Granovetter, 1973). In contrast, strong ties embody relationships with greater emotional intimacy, support, and more frequent interactions (Granovetter, 1973), which can provide emotional support and reinforce tight-knit group identities (Putnam, 2001).

Whereas some social media sites reflect symmetric and reciprocal ties, such as Facebook "friends," (Ellison et al., 2007), SLSSs mostly reflect one-directional "follow" relationships with users able to "follow" streamers, which enable notifications whenever the streamer goes live, and can "subscribe" to streamers for a monthly fee for a membership badge and custom items as a that encourage and financially support streamers' channel. These relationships are largely one-directional, as streamers can have thousands of viewers and receive many comments but rarely reciprocate the attention and engagement at the same degree (Ding et al., 2012).

Despite their asymmetric weak tie network composition, SLSSs are "rich" media, which afford immediate feedback with multiple audio and video cues, channels, and languages that can enable more immediate, efficient and clear communication between streamers and viewers (Hsu, Lin, & Miao, 2020). Immediacy and richness of information fosters a more intimate channel that is especially valued in more close relationships (Liu et al., 2019). Indeed, individuals often select certain media for certain relationships, with rich media often used with strong ties, and media that communicate fewer cues used with weak ties (Baym, Zhang, & Lin, 2004; Goodman-Deane et al., 2016). With both video and text-chat that afford more social cues and synchronous communication, viewers could experience heightened intimacy and perceived relational closeness with a streamer. In addition, SLSSs also afford a unique means of participation by allowing viewers to make direct financial contributions to the streamer as a means of long-term financial and emotional support (Johnson & Woodcock, 2019b). Therefore, viewers can gain visibility and build connections not just by social interactions, but also by financial contributions, which may have downstream consequences for their well-being.

**SLSSs Use & Psychological Well-being**

Psychological well-being (PWB) reflects optimal psychological health and functioning based on an individual's positive relationships, sense of life purpose, self-acceptance, personal growth, autonomy, and mastery (Ryff & Keyes, 1995). Recent research has highlighted the differential well-being outcomes from specific online activities (Burke & Kraut, 2016; Ernala et al., 2020), with some research examining active versus passive use (Verduyn et al., 2015; Escobar-Viera et al., 2018), directed communication versus consumption (Burke, Marlow, & Lento, 2010; Burke & Kraut, 2016), or degree of sociality (Gerson, Plagnol, & Corr, 2017). One study found that receiving targeted and personalized communication such as a comment or message from strong ties were associated with improvements in well-being while receiving one-click interactions such as "likes" from weak ties and viewing broadcasting content (even with strong ties) were not (Burke & Kraut, 2016), suggesting that active use to connect with strong ties leads to well-being benefits by increasing perceptions of social support.

Active use encompasses activities that leave visible traces and facilitate direct interactions with others via commenting and posting (Verduyn et al., 2015; Verduyn et al., 2017). Within SLSSs, viewers can actively participate by commenting in a streamer's chatroom or "Chat," responding to other commenters, and by spamming "Emotes," which are custom streamer emojis imbued with meaning. In their meta-analysis, Meier & Reinecke (2020) found that interactions (replying, commenting, and liking) were positively related to overall well-being, likely due to receiving emotional support from friends and acquaintances which can provide benefits to well-being (Burke et al., 2010).

*H1. Active participation is positively related to PWB.*

Paying for premium services is strongly associated with active social behaviors in online communities; subscribers tend to have more connections and are often more participatory, with their payment contributing to an improved content experience (Oestreicher-Singer & Zalmanson, 2009). Research indicates that spending money on others may increase happiness and potentially enhance well-being (Dunn, Aknin, & Norton, 2014; Diener, Lucas, & Oishi, 2018). In SLSSs, viewers can engage in a more committed form of active participation by financially contributing directly to a streamer via monthly subscriptions or donations. Subscribers and donors can attach a custom message to their contribution that is highly visible on stream and often elicits a direct response from the streamer in-real-time. Financial contributions therefore provide a means to directly socialize and self-disclose with a streamer and reflect more effortful active participation beyond simply posting in Chat (Johnson & Woodcock, 2019b; Wohn et al., 2018). Tamir and Mitchell (2012) demonstrated that individuals are willing to forgo money for the opportunity to disclose information about themselves to others which may lead to a response, liking, and possibly stronger social attachments and relationships. This act of self-disclosure, independent of social interaction, may itself promote psychological well-being (Pennebaker & Chung, 2011). Streamers' response to financial contributors may further elicit their feelings of trust, commitment, and closeness. We therefore predict:

*H2. Financial commitment is positively related to PWB.*

Passive consumption is the most common activity on social media sites (Krasnova et al., 2013) and is related to decreased well-being (Meier & Reinecke, 2020; Verduyn et al., 2015). Passive use involves the monitoring or consumption of content without interactions and can include scrolling through a newsfeed or simply viewing content (Verduyn et al., 2017). In SLSSs, passive participation consists of watching a channel without further engagement and information search behaviors such as looking for streams to watch (Bründl, 2018). SLSS viewers often switch from passive to active engagement, with many engaging in both behavior types (Ask, Spilker, & Hansen, 2019).

There is robust support for the negative association between passive participation and PWB, often explained by social comparison and envy (Verduyn et al., 2017; Liu et al., 2019). However, the displacement hypothesis may be more appropriate for SLSS contexts. The displacement hypothesis predicts a negative association between time spent online and well-being; as time is inelastic, investing time in online spaces detracts or displaces time that would otherwise be spent investing in offline activities and relationships, leaving individuals feeling lonelier (Nie, 2001). Research has supported this, with greater Internet use and time spent passively browsing social media associated with decreased social involvement and psychological well-being, (Kraut et al., 1998; Verduyn et al., 2015; Deters & Mehl, 2012; Joseph, 2020). One

LIVE-STREAMING USE, SOCIAL CAPITAL & WELL-BEING                                9study found that Facebook activity led to a deterioration of mood due to the perception of having wasted time (Sagioglou & Greitemeyer, 2014). Therefore:

*H3. Passive participation is negatively related to PWB.*

**The Mediating Role of Social Capital**

As the relationship between participation and PWB is supported by robust empirical data, scholars have suggested that social capital as an important explanatory mechanism (see Spottswood & Wohn, 2020 for a review). Social capital is the sum of actual and potential resources embedded within and accessed through an individual's network of relationships (Nahapiet & Ghoshal, 1998). Social capital can take multiple forms, but always consists of a social structure and the individual actions taken within that structure (Coleman, 1988; Putnam, 2001). As individuals interact and strengthen their social relationships, social capital forms and creates value for the community (Arregle et al., 2007), embodying the multiple resources that are derived from social relationships, such as norms of reciprocity and values systems (Tsai & Ghoshal, 1998; Putnam, 2001; Shen, Monge, & Williams, 2014).

According to Nahapiet and Ghoshal (1998), social capital encompasses three distinct dimensions that capture the structural, relational, and cognitive aspects of social resources accessed from social ties. Structural social capital considers the social ties or connections facilitated by interactions that create network structure and provide individuals access to information, knowledge, and other social resources. Many scholars have distinguished these ties as bonding and bridging social capital (Williams, 2006; Claridge, 2018). Relational social capital captures the quality of those relationships. As individuals interact over time, their relationships may deepen to reflect trust, norms and sanction, obligations and expectations, and identity and identification that comprise the relational dimension (Nahapiet & Ghoshal, 1998). Relational social capital encourages normative group behavior and relational cohesion based on trust, reciprocity, and a shared identity but is distinct from these concepts individually. At its core, relational social capital captures associability, or the willingness to prioritize collective goals over individual goals (Lazarova & Taylor, 2009; Claridge, 2018). Lastly, cognitive social capital describes the wider social context manifested in shared representations, interpretations, and meaning with a group (Nahapiet & Ghoshal, 1998). Cognitive social capital captures systems of meaning generated by communities, including vocabulary and shared goals, vision, and values that allow a common understanding of community norms (Nahapiet & Ghosal, 1998).

 Despite their distinct definitions, the three dimensions of social capital are highly interconnected (Claridge, 2018). Social interactions and resulting connections are required for the development of relational and cognitive social capital which may further reinforce and develop structural social capital by providing common ground and mutual trust and identification which may motivate interactions and the formation of new relationships (Claridge, 2018). Indeed, making new connections is a common motivator for participation across online communities and SLSSs (Brandtzaeg & Heim, 2009; Bründl & Hess, 2016; Hilvert-Bruce et al., 2018). As individuals actively participate in a streamers' Chat, they are likely to develop more social connections, or structural social capital within that community. Individuals' participation and shared experiences within the community may further foster a sense of community and group identity, captured by relational social capital (Hilvert-Bruce et al., 2018; Van den Hoof & Huysman, 2009). Chat in SLSSs often employ short-hand communication and Emotes that are often used in response to certain events; this shared language and meaning, or cognitive social capital, may distinguish ingroup fans or members from casual spectators who may not understand (Carter & Egliston, 2018). Considering how close relationships are imperative to well-being (Valkenburg & Peter, 2007), more social interaction ties, shared community norms and identity, as well as common language and values may satisfy individuals' inherent need to belong which may positively affect their PWB (Baumeister & Leary, 1995).

*H4. a) Structural social capital, b) relational social capital, and c) cognitive social capital positively mediates the relationship between active participation and PWB.*

Among active participants, Subscribers tend to have more friends and connections to a community; they tend to post more, lead more groups, and create more content (Oestreicher-Singer & Zalmanson, 2009). Increased commitment and involvement may translate into social connections, or structural social capital, and help users gain visibility and access to other social resources (Coleman, 1988). Additionally, subscribers and donors can send a custom message to the streamer with their financial contribution; this direct communication may reflect a closer attachment to the streamer, as well as greater shared identification and values with the streamer, thereby relating to an increase in relational and cognitive social capital. Financial contributions can materialize as a form of emotional support to the streamer (Wohn et al., 2018); as viewers become more invested in their favorite streamer, they may desire to financially reciprocate (Diwanji et al., 2020). Payments to SLSS streamers often accompany premium and exclusive content ranging from specialized Emotes, subscriber-only Chats, ad-free viewing experiences, and badges that signal relational investment. Financial commitment may help members build their reputation and garner the streamer's attention for a potential promotion to a community moderator (Wohn, 2019). Therefore, paying members may develop greater social capital, supporting a streamer's vision for their community and acting as informal and formal moderators to ensure the language, goals, and norms of the community are maintained. Therefore:

*H5. a) Structural social capital, b) relational social capital, and c) cognitive social capital positively mediates the relationship between financial commitment and PWB.*

As passive viewers monitor or consume content without interactions (Verduyn et al., 2017), they may not have the same access to the social connections or support of active or paying viewers. However, even without direct or active participation, individuals may still experience gains from their passive use. Passive participation is still participation in a limited capacity that is more active than not tuning in. In witnessing the interactions between the streamer and other viewers, passive participants may experience emotional connectedness and engagement from the fast-moving Chat and social interactions amongst members (Shin, 2019; Lim et al., 2020). Some studies suggest that passive members do not feel alienated or disconnected from their social institutions and may still reap benefits that are embedded within their groups (Wollebaek & Selle, 2003). Lakey and colleagues (2014) found that merely observing other people's conversations and activities may elicit positive affect and perceived support. These passive members may still feel a sense of community and mutual identification by watching others which may aid in the development of their relational social capital. A recent study found that SLSS viewers that experienced flow, or the act of total concentration and enjoyment, reported greater satisfaction with their experience. In turn, satisfaction enhanced respondents' social well-being and decreased feelings of loneliness (Kim & Kim, 2020). Passive spectators may experience immersion, which may aid in their socialization of shared community representations, language, and values, or cognitive social capital. As passive viewers observe, they slowly acculturate themselves and gather discursive tools which may eventually motivate them toward more visible and active participation (Georgen et al., 2015). In this way, passive viewers may still experience benefits in relational social capital and cognitive social capital.

*RQ1. How do a) structural social capital, b) relational social capital and c) cognitive social capital mediate the relationship between passive participation and PWB?*

**The Role of Parasocial Relationship**

In addition to social capital, parasocial relationship could also explain the association between SLSS participation and PWB. Parasocial relationship (PSR) is an enduring asymmetrical relationship based on repeated encounters during which a user forms a socioemotional bond with a media performer (Horton & Wohl, 1956; Dibble, Hartmann, & Rosaen, 2016). A media performer may encourage PSR development by using conversational and informal communication and gestures that emulates interpersonal communication in face-to-face settings (Horton & Wohl, 1956). Over time, users may perceive the media personality as directly interacting with them which in turn may increase emotional bonding and the feeling of intimately "knowing" the media personality (Ding et al., 2012). Live-streamers often interact directly with their viewers in real time which may facilitate the development of PSR (Hou, Guan, Li, & Chong, 2019). PSR is related to media enjoyment (Wulf et al., 2020), intention to continue viewing content (Hu et al., 2017; Lim et al., 2020), and willingness to financially support a streamer (Wohn et al., 2018).

Research examining PSR and PWB is scarce with conflicting findings (Hartman, 2016). The asymmetric nature of PSR may hurt social capital, increase loneliness, and may be indicative of social media addiction (Burke et al., 2010; Baek, Bae, & Jang, 2013). According to the compensation hypothesis, individuals who lack substantial social relationships may seek out asymmetric ones to compensate (Hartman, 2016). On the other hand, PSR may be positively related to PWB by satisfying one's need to belong (Hartmann, 2016), which can occur when social interactions are pleasant and characterized by stability, mutual concern, and intention to continue a relationship (Baumeister & Leary, 1995). If satisfied, the need to belong promotes subjective well-being (Hartmann, 2016). Interactions with a mediated other may trigger enjoyment or the perception of social support which improves well-being (Lakey et al., 2014).

*RQ2. Does PSR mediate the relationship between a) active participation and PWB, b) passive participation and PWB, and c) financial commitment and PWB.*

## Method

**Twitch Context**

Twitch.tv is one of the most popular SLSSs, boasting 9.7 million unique streamers each month and an average of 3 million concurrent viewers at any given time (Twitch Statistics & Charts, 2021). With its focus on amateur broadcasting, anyone with a free account can publicly broadcast live audio and video to small and large audiences. Essential to the live-streaming and Twitch experience, viewers can interact with the streamer and one another in real time through Chat. Viewers can also "follow" a streamer to receive notifications every time the streamer goes live, "subscribe" to a streamer at a monthly fee ($4.99, $9.99, or $24.99) for custom Emotes and badges that indicate their membership status, "gift" subscriptions to other viewers, and lastly "cheer" for a streamer by donating bits, which is a Twitch currency where 1 bit is the equivalent of 1 cent. As streamers interact with their audiences, establish relationships with viewers, and grow their communities, some eventually turn to full-time streaming where they earn a reliable income through viewers' subscriptions, gifts, and bits.

**Measures**

  **Psychological well-being.** PWB was measured using the Brief Inventory of Thriving (BIT) (Su, Tay, & Diener, 2014). Created to synthesize the well-being literature, BIT captures all the core subdimensions of positive psychological health and functioning including subjective well-being or the feeling of satisfaction, supportive positive relationships, interest in daily activities, sense of accomplishment, autonomy, meaning in life, and optimism (Su et al., 2014). Ten items measured participants' PWB with a 5-point Likert scale where 1 = strongly disagree and 5 = strongly agree. Sample items include, "I feel a sense of belonging in my community," "I can succeed if I put my mind to it," and "I am optimistic about my future" (α = .93).

**Social capital.** All three dimensions of social capital were measured using existing scales (Chiu, Hsu, & Wang, 2006; Lin & Lu, 2011) on 5-point Likert scale (1= strongly disagree; 5 = strongly agree). Structural social capital was measured with four items, such as "I maintain close social relationships with other Twitch users who watch {favorite streamer}'s channel," and "I have frequent communication with some Twitch users who watch {favorite streamer}'s channel." (α = .908). Relational social capital was measured using four items such as "I feel a sense of belonging toward {favorite streamer}'s Twitch channel," and "I am proud to be a viewer of {favorite streamer}'s Twitch channel." (α = .848). Cognitive social capital was measured using six items such as, "{Favorite streamer}'s Twitch channel activities are in line with my personal values," and "I agree with what {favorite streamer}'s Twitch channel considers to be important" (α = .88).

**Parasocial relationship.** Thirteen items measured participants' PSR with their favorite streamer, using Wulf and colleagues' (2020) adaptation of Hartmann and colleagues' (2008) positive PSR scale for Twitch, on a 5-point Likert scale (1 = "strongly disagree" and 5 = "strongly agree"). The scale measures the intimacy of the perceived relationship as well as viewer's interest in the media figure. Sample items include, "My favorite Twitch streamer makes me feel as comfortable as when I am with friends," and "I would like to meet my favorite Twitch streamer in person" (α = .87).

**Participation.** We adapted the Passive Active Use Measure (PAUM; Gerson, Plagnol & Corr, 2017), originally developed to categorize various Facebook activities as active or passive use, to distinguish various Twitch activities and behaviors into active or passive behaviors while watching respondents' favorite streamer. Active participation was measured by taking the average score from items that included "Commenting directly to the streamer," "Responding or Reacting," and "Spamming Emotes" which were answered on a 5-point Likert scale ranging from "Never (0% of the time)" to "Very frequently (100% of the time)" ($\alpha = .90$). Passive participation item included, "Reading Twitch Chat."

**Financial commitment.** Respondents were asked whether they had ever subscribed or are currently subscribed to their favorite streamer. If they responded yes, they were asked for how many months and at what Tier plan, which was used to calculate a total subscription amount. Similarly, respondents were asked whether they had donated to their favorite streamer and were asked to estimate the dollar amount. Financial commitment was then measured by taking the sum of both subscription and donations.

**Control Variables.** We controlled for the amount of time (in hours) respondents spent watching their favorite streamer in the last week. Perceived offline social support was measured using the 12-item Interpersonal Support Evaluation List (ISEL-12) (Cohen et al., 1985) ($\alpha = .90$). Additional variables included respondents' education, sex, and a measure of the extent that health issues (mental and physical) affected respondents' daily life for over a week (from "Never" to "Frequently"), all of which may affect users' well-being.

**Participants and Procedure**

We defined eligible participants as those who were 18+ years of age and had a Twitch account active in the past month. We first sent out a screening survey to 980 participants over

Prolific.co, an online participant pool, and 665 respondents (68%) met the eligibility criteria. We then invited them to complete the full online survey, and 427 (64%) completed in late February 2021. Participants first reported their PWB and perceived offline social support, and then were asked questions about PSR, social capital, time spent watching their favorite streamer in the past week, the passive and active behaviors, subscription/donation, as well as information regarding their account age and number of streamers they follow. Lastly, respondents were asked questions on their demographic information and health status. To ensure data integrity, we included an attention check question in the survey, and also manually checked whether the respondent's favorite streamer mentioned existed on Twitch. 21 responses were excluded from the analysis; 14 participants failed the attention check and 7 participants responded with false streamer information, resulting in a sample size of 396. Participants were compensated for both the screening survey (1 minute) and the full survey (12 minutes) with an hourly rate of $14.50.

68% of participants identified as male, 28% as female, 3% as non-binary, and 1% preferred not to identify. Respondents spent 9.59 hours on average watching their favorite Twitch streamer per week (*Median* = 5, *SD* = 11.75), had a Twitch account for over 4 years (*Median* = 4, *SD* = 2.39) and contributed $21.57 to their favorite streamer in total since first subscribing or following (*Median* = 4.99, *SD* = 45.32) (see Table 1).

**Analysis**

Structural Equation Modeling (SEM) was used to analyze the data, using the *lavaan* package (Rosseel, 2012) in *R* and bias-corrected 95% bootstrapped confidence intervals (CI) based on 5000 resamples. We ran two structural equation models for each proposed mediator; Model 1 was the baseline model (see Fig. 1, 2), and Model 2 added control variables including time spent, perceived social support, education, health, and sex.[i] As both Model 1 and Model 2 are fully saturated, meaning all variances and covariances of the variables are estimated as model parameters, goodness of fit indicators such as Comparative Fit Index (CFI), Root Mean Square Error of Approximation (RMSEA), and Standardized Root Mean Square Residual (SRMR) were equivalent across models. Goodness of fit was therefore assessed by comparing the change in R-Squared. Descriptive statistics are presented in Table 1.

For the social capital models, Model 2 with added covariates explained 43% more variance in PWB, 2% more variance in structural social capital, 6% more variance in relational social capital, and 4% more variance in cognitive social capital compared to Model 1. For the PSR models, Model 2 with covariates explained almost 38% more variance in PWB than Model 1, and 1% more variance in PSR. We therefore report Model 2 results for social capital and PSR to address our hypotheses and research questions.

## Results

H1 predicted active participation will be positively related to PWB. This relationship was significant ($\beta$ [bootstrap 95% CI] = .089 [.003, .174], $Z$ = 2.028), supporting the direct effect of active participation onto psychological well-being. H2 similarly predicted a positive direct relationship between financial commitment and PWB, however, this was not significant ($\beta$ [bootstrap 95% CI] = 0 [-.192, .152], $Z$ = .005). H3 predicted a negative direct relationship between passive participation and PWB, however, this was also not significant ($\beta$ [bootstrap 95% CI] = -.01 [-.08, .057], $Z$ = -.291).

H4 predicted the three dimensions of social capital will positively mediate the relationship between active participation and PWB. Structural social capital was a significant mediator ($\beta$ [bootstrap 95% CI] = .045 [.012, .089], $Z$ = 2.273) between active participation and PWB, but relational social capital ($\beta$ [bootstrap 95% CI] = .013 [-.009, .038], $Z$ = 1.104) and cognitive social capital (β [bootstrap 95% CI] = -.011 [-.033, .000], Z = -1.333) were not significant mediators.

H5 predicted the three dimensions of social capital will positively mediate the relationship between financial commitment and PWB. Structural social capital was a significant mediator (β [bootstrap 95% CI] = .023 [.002, .070], Z = 1.456), but relational social capital (β [bootstrap 95% CI] = .014 [-.008, .050], Z = 0.985) and cognitive social capital (β [bootstrap 95% CI] = -.017 [-.052, .000], Z = -1.316) did not significantly mediate the relationship between financial commitment and PWB.

RQ1 examined how a) structural social capital, b) relational social capital and c) cognitive social capital mediate the relationship between passive participation and PWB. Structural social capital (β [bootstrap 95% CI] = -.006 [-.023, .002], Z = -1.058), relational social capital (β [bootstrap 95% CI] = .006 [-.002, .023], Z = .994) and cognitive social capital (β [bootstrap 95% CI] = -.007 [-.026, .001], Z = 2.273) did not mediate the relationship between passive participation and PWB.

RQ2a, b, and c examined whether PSR mediated the relationship between Twitch use and PWB. While active participation was significantly related to PSR (β [bootstrap 95% CI] = .18 [.102, .259], Z = 4.485), PSR did not mediate the relationship between active participation and PWB (β [bootstrap 95% CI] = .006 [-.012, .027], Z = .594), financial commitment and PWB (β [bootstrap 95% CI] = .002 [-.003, .014], Z = .51), nor passive participation and PWB (β [bootstrap 95% CI] = .005 [-.008, .032], Z = .5).

## Discussion

Examining the potential mediating role of social capital and PSR, this study examines how participation in the popular SLSS, Twitch.tv, is related to viewers' psychological well-being. Findings demonstrate that active participants reported significantly greater PWB, partially explained by their structural social capital, or social connections within the community. Individuals who financially contributed more to their favorite streamer also reaped PWB benefits by way of their social ties. Interestingly, individuals who reported having greater cognitive social capital, or shared values with their favorite streamer, experienced lower PWB. Ultimately, while structural social capital explained the relationship between active use and positive PWB, and financial commitment and PWB, relational social capital which captured shared identity with a streamer and parasocial relationship which captured the perceived intimacy between individuals and their favorite streamer were not instrumental to participants' PWB. Examining the broad genres on social live streaming services beyond gaming, this study contributes to the field by demonstrating the value of social connections in providing psychological benefits to SLSS users and the benefits active and committed use have on PWB.

**Active Use Promotes PWB Directly and Via SLSS Social Ties**

*"Rich Get Richer" Hypothesis*

In line with prior research in social media contexts, active SLSS use, rather than passive use, was positively related to PWB (Verduyn et al., 2017; Meier & Reinecke, 2020), after controlling for demographic variables, perceived offline social support, and health status. However, it is unclear whether the positive benefits of active SLSS participation is conditional on users' existing psychological well-being, sometimes referred to as the "rich-get-richer" phenomenon in earlier work on Internet sociality (Kraut et al., 2002). One study found that adolescents who already had strong social relationships at early ages were more likely to use online communication which in turn predicted more cohesive friendships and better connectedness (Lee, 2009). In other words, it is possible that individuals who actively participate in Chat already have higher PWB to begin with. Viewers may have robust offline social networks that provide social support and supplement these social interactions with SLSS activities, reflecting Bekalu and colleagues (2019) findings that SLSS integration into social routines is related to greater well-being, mental health, and health outcomes.

The "rich-get-richer" hypothesis may be further supported by our findings that structural social capital positively mediates the relationship between active participation and PWB, as well as between financial commitment and PWB. As users are motivated to use SLSSs for their interactivity and increased sociability (Hilvert-Bruce et al., 2018), active or committed actions taken during individuals' favorite streamers' live-sessions may facilitate stronger connections. Individuals with higher PWB may also have existing offline friendships that are also present and actively involved in a favorite streamer's channel, with our findings demonstrating perceived offline social support was positively correlated to structural social capital, relational social capital, and PWB. Similarly, individuals with more disposable income that enables them to financially contribute more may have higher PWB to begin with. Prior research suggests a positive association between household income and subjective well-being, with income consistently associated with life evaluations over time (Diener, Tay, & Oishi, 2013). Future studies should consider including disposable income and leisure time as covariates.

*"Social Compensation" Hypothesis*

Our findings do not rule out the "social compensation" hypothesis, which asserts that the Internet may be more beneficial for socially isolated individuals who compensate for a lack of offline social support by developing relationships online (McKenna & Bargh, 1999). Actively participating in SLSS Chat may serve a self-affirming purpose for viewers (Toma & Hancock, 2013), with the immediate nature of Chat feedback facilitating the receival of emotional rewards or online social support (Reinecke & Trepte, 2014) that may yield PWB benefits by fulfilling individuals' personal integrative or self-presentation needs (Hsu, Tien, Lin, & Chang, 2015). Future research should examine both the rich-get-richer and social compensation hypothesis.

**PSR and Cognitive Social Capital: Displaced Investments**

*"Displacement Hypothesis"*

Our findings fail to demonstrate the connection between relational and cognitive social capital with PWB. However, while the cognitive dimension was not a significant mediator, we did find that greater cognitive social capital was related to decreased PWB. This could be explained by the displacement hypothesis which asserts that time spent online may displace valuable time interacting with offline and more socially rewarding connections (Kraut et al., 1998), which may lead to an overall PWB deficit. This is not to say that SLSS relationships cannot yield PWB benefits, but rather that the strength of these relationships and the social support received from them may not compare to offline or non-SLSS online relationships. Individuals who reported stronger alignment with a streamer's values and vision may be displacing emotional investment and time away from their own personal life and that the shared language and values from the SLSS community have different or no meaning outside of the community (Ansari et al., 2012). In fact, time spent emotionally investing in a streamers' community values and goals may be indicative of problematic use and dependency. A recent survey of American adults found that individuals with greater reliance on social media reported lower social well-being, positive mental health, and self-rated health outcomes (Bekalu et al., 2019). Some researchers suggest that sole reliance on social media to relieve stress, loneliness, or depression for individuals with poorer well-being may be a precursor to problematic use (Xu & Tan, 2012). While longer exposure to a favorite streamers' channel enculturates individuals' to the norms, meanings, and goals of that community, it may take time away from individuals' investment into their own values and goals or reflect individuals with problematic dependency. This may also be the case with PSR with a streamer which, while insignificant in our study, may have a curvilinear effect on viewers' behavior and well-being. As viewers spend more time developing the perception of intimacy with a streamer, they may displace time they could have spent developing a personal and symmetric relationship. This dependency on an asymmetric parasocial relationship with a streamer may lead to greater loneliness and social media addiction (Baek et al., 2013). Considering its prevalent role in SSLS use, researchers should further examine the benefits PSR and relational and cognitive social capital may have on individuals and communities. Future research should also explore individuals' pre-existing mental health and their reliance on SLSSs to understand potential tradeoffs different dimensions of social capital have for various populations' PWB.

**Limitations & Future Directions**

This most significant limitation of the current study is the cross-sectional design and lack of causality. As discussed above, it is possible that individuals with high PWB actively participate and financially contribute more than individuals with lower PWB. Second, survey data was collected during the COVID-19 pandemic at the start of vaccine distribution, which may have influenced participants' viewing and participatory behaviors, attachment to a streamer and wider community, perceived offline support, and financial gifting behaviors. A replication could validate the potential temporal effect of COVID-19 and establish a more direct link between SLSS use and its effect on users' psychological well-being. Third, while this study examined both social capital and PSR as potential mechanisms, other theories may apply, such as the feeling of having wasted time (Sagioglou & Greitemeyer, 2014). Additionally, our study examined a relatively small sample which may introduce bias in estimation accuracy and model specification. Therefore, future studies should examine larger samples of SLSS users across live-streaming platforms so that the benefits or downsides of SLSS on well-being can be further established. Future research can also distinguish directed and targeted communication, broadcast communication, and one-click communication to examine how different actions with different audiences relate to PWB. Finally, future research could employ a variety of mixed-methods, such as survey data combined with behavioral data to cross-validate findings and ascertain how use benefits individuals' well-being. Future research should employ longitudinal analysis to better ascertain directions of causality.

**Conclusion**

Similar to other social media platforms, SLSSs should not be treated as a black box or a monolithic source of psychological effects. Our findings illustrate how individual use and interactions with streamers and other viewers may result in different PWB outcomes, with active participation positively related to higher PWB. The distinction between active and passive participation allows for greater nuance and generalizability over time and comparisons across platforms, and our examination of financial commitment demonstrates how this unique SLSS means of interaction may relate to users' PWB, which few studies have addressed. In using a neutral indicator of PWB rather than more commonly used negative indicators such as depression and loneliness (Chen & Chang, 2019; Wan & Wu, 2020), our study avoids a negativity bias, and highlights the potential positive outcomes that can arise from interactive SLSS use. Lastly, rather than narrowly focusing on e-Sports or video game live-streaming communities only, our study surveyed respondents across content categories for a more comprehensive understanding of SLSS users. As one of the first studies to examine the role of social capital and parasocial relationships on SLSS viewers' psychological well-being, our findings indicate not all forms of social capital are equally beneficial to individuals' PWB. Structural social capital which captures the number and strength of social relationships in a SLSS community relate to higher PWB for individuals who actively participate in Chat and financially contribute to their favorite streamer, potentially highlighting the role of the "rich-get-richer" hypothesis. In contrast, cognitive social capital, or sharing in the values and goals of the streamer and their community and having a strong parasocial relationship with a streamer may displace time away from viewers' shared values and goals with offline or symmetric social relationships, which may relate to lower PWB. As SLSSs continue to grow in popularity and platforms afford additional social interactions through financial contributions, scholars should continue to assess how and with whom users interact with to better understand its relationship to individuals' psychological well-being.

**Availability of Data**

The data underlying this article will be shared on reasonable request to the corresponding author.

LIVE-STREAMING USE, SOCIAL CAPITAL & WELL-BEING                27*Communities and Social Computing* (pp. 143-152). Springer, Berlin, Heidelberg. https://doi.org/10.1007/978-3-642-02774-1_16

Bründl, S. (2018, January). Passive, Active, or Co-Active? The Link Between Synchronous User Participation and Willingness to Pay for Premium Options. *Proceedings of the 51st Hawaii International Conference on System Sciences, USA*. 483-492. https://doi.org/10.24251/hicss.2018.063

Bründl, S., & Hess, T. (2016, June). Why do Users Broadcast? Examining Individual Motives and Social Capital on Social Live Streaming Platforms. *Proceedings of the Pacific Asian Conference on Information Systems, Taiwan*. 332. https://aisel.aisnet.org/pacis2016/332/

Burke, M., & Kraut, R. E. (2016). The relationship between Facebook use and well-being depends on communication type and tie strength. *Journal of Computer-Mediated Communication*, *21*(4), 265-281. https://doi.org/10.1111/jcc4.12162

Burke, M., Marlow, C., & Lento, T. (2010, April). Social network activity and social well-being. *Proceedings of the SIGCHI Conference on Human Factors in Computing Systems, USA*, 1909-1912. https://doi.org/10.1145/1753326.1753613

Carter, M., & Egliston, B. (2018). Audiencing on Twitch. *Proceedings from Digital Games Research Association Annual Conference*, *Italy. http://www.digra.org/wp-content/uploads/digital-library/DIGRA_2018_paper_167.pdf*

Chen, C. Y., & Chang, S. L. (2019). Moderating effects of information-oriented versus escapism-oriented motivations on the relationship between psychological well-being and problematic use of video game live-streaming services. *Journal of Behavioral Addictions*, *8*(3), 564-573. https://doi.org/10.1556/2006.8.2019.34

LIVE-STREAMING USE, SOCIAL CAPITAL & WELL-BEING                                                              32Johnson, M. R., & Woodcock, J. (2019a). The impacts of live streaming and Twitch.tv on the video game industry. *Media, Culture & Society*, *41*(5), 670–688. https://doi.org/10.1177/0163443718818363

Johnson, M. R., & Woodcock, J. (2019b). "And today's top donator is": How live streamers on Twitch. tv monetize and gamify their broadcasts. *Social Media+ Society*, *5*(4), 1-11. https://doi.org/10.1177/2056305119881694

Joseph, J. J. (2020). Facebook, Social Comparison, and Subjective Well-Being: An Examination of the Interaction Between Active and Passive Facebook Use on Subjective Well-Being. In M. Desjarlais (Ed.), *The Psychology and Dynamics Behind Social Media Interactions* (pp. 268-288). IGI Global. https://doi.org/10.4018/978-1-5225-9412-3.ch011

Kim, H. S., & Kim, M. (2020). Viewing sports online together? Psychological consequences on social live streaming service usage. *Sport Management Review*, *23*(5), 869-882. https://doi.org/10.1016/j.smr.2019.12.007

Kim, C., & Shen, C. (2020). Connecting activities on Social Network Sites and life satisfaction: A comparison of older and younger users. *Computers in Human Behavior*, *105*, 106222. https://doi.org/10.1016/j.chb.2019.106222

Krasnova, H., Wenninger, H., Widjaja, T., & Buxmann, P. (2013). Envy on Facebook: a hidden threat to users' life satisfaction? *Proceedings of the 11th International Conference on* Wirtschaftsinformatik, Germany. https://doi.org/ 10.7892/BORIS.47080

Kraut, R., Patterson, M., Lundmark, V., Kiesler, S., Mukophadhyay, T., & Scherlis, W. (1998). Internet paradox: A social technology that reduces social involvement and psychological well-being?. *American Psychologist*, *53*(9), 1017-1031. https://doi.org/10.1037/0003-066X.53.9.1017

LIVE-STREAMING USE, SOCIAL CAPITAL & WELL-BEING                                    34Meier, A., & Reinecke, L. (2020). Computer-mediated communication, social media, and mental health: A conceptual and empirical meta-review. *Communication Research*, *48*(8), 1182-1209. https://doi.org/10.1177/0093650220958224

Nahapiet, J., & Ghoshal, S. (1998). Social capital, intellectual capital, and the organizational advantage. *Academy of Management Review*, *23*(2), 242-266. https://doi.org/10.2307/259373

Nie, N. H. (2001). Sociability, interpersonal relations, and the Internet: Reconciling conflicting findings. *American Behavioral Scientist*, *45*(3), 420-435. https://doi.org/10.1177/00027640121957277

Oestreicher-Singer, G., & Zalmanson, L. (2009). " Paying for Content or Paying for Community?" The Effect of Social Involvement on Subscribing to Media Web Sites. *International Conference on Information System Proceedings*, USA. https://aisel.aisnet.org/icis2009/9

Pennebaker, J. W., & Chung, C. K. (2011). *Expressive writing: Connections to physical and mental health.* In H. S. Friedman (Ed.), *Oxford Library of Psychology. The Oxford Handbook of Health Psychology* (pp. 417–437). Oxford University Press. https://doi.org/10.1093/oxfordhb/9780195342819.013.0018

Putnam, R. (2001). Social capital: Measurement and consequences. *Canadian Journal of Policy Research*, *2*(1), 41-51. https://smg.media.mit.edu/library/putnam.pdf

Reinecke, L., & Trepte, S. (2014). Authenticity and well-being on social network sites: A two-wave longitudinal study on the effects of online authenticity and the positivity bias in SNS communication. *Computers in Human Behavior*, *30*, 95-102. https://doi.org/10.1016/j.chb.2013.07.030

LIVE-STREAMING USE, SOCIAL CAPITAL & WELL-BEING                                                                 36Toma, C. L., & Hancock, J. T. (2013). Self-affirmation underlies Facebook use. *Personality and Social Psychology Bulletin*, *39*(3), 321-331. https://doi.org/10.1177/0146167212474694

Tsai, W., & Ghoshal, S. (1998). Social capital and value creation: The role of intrafirm networks. *Academy of Management Journal*, *41*(4), 464-476. https://doi.org/10.2307/257085

Valkenburg, P. M., & Peter, J. (2007). Online communication and adolescent well-being: Testing the stimulation versus the displacement hypothesis. *Journal of Computer-Mediated Communication*, 12(4), 1169-1182. https://doi.org/10.1111/j.1083-6101.2007.00368.x

Verduyn, P., Lee, D. S., Park, J., Shablack, H., Orvell, A., Bayer, J., Ybarra, O., Jonides, J., & Kross, E. (2015). Passive Facebook usage undermines affective well-being: Experimental and longitudinal evidence. *Journal of Experimental Psychology: General*, *144*(2), 480. https://doi.org/10.1037/xge0000057

Verduyn, P., Ybarra, O., Résibois, M., Jonides, J., & Kross, E. (2017). Do social network sites enhance or undermine subjective well-being? A critical review. *Social Issues and Policy Review*, *11*(1), 274-302. https://doi.org/10.1111/sipr.12033

Wan, A., & Wu, L. (2020). Understanding the Negative Consequences of Watching Social Live Streaming Among Chinese Viewers. *International Journal of Communication*, *14*(2020), 5311-5330. https://ijoc.org/index.php/ijoc/article/view/13294/3249

Williams, D. (2006). On and off the 'Net: Scales for social capital in an online era. *Journal of Computer-Mediated Communication*, *11*(2), 593-628. https://doi.org/10.1111/j.1083-6101.2006.00029.x

Wohn, D. Y. (2019, May). Volunteer moderators in twitch micro communities: How they get involved, the roles they play, and the emotional labor they experience. In *Proceedings of*

Figure 1. SEM with 3 Dimensions of Social Capital as Mediators

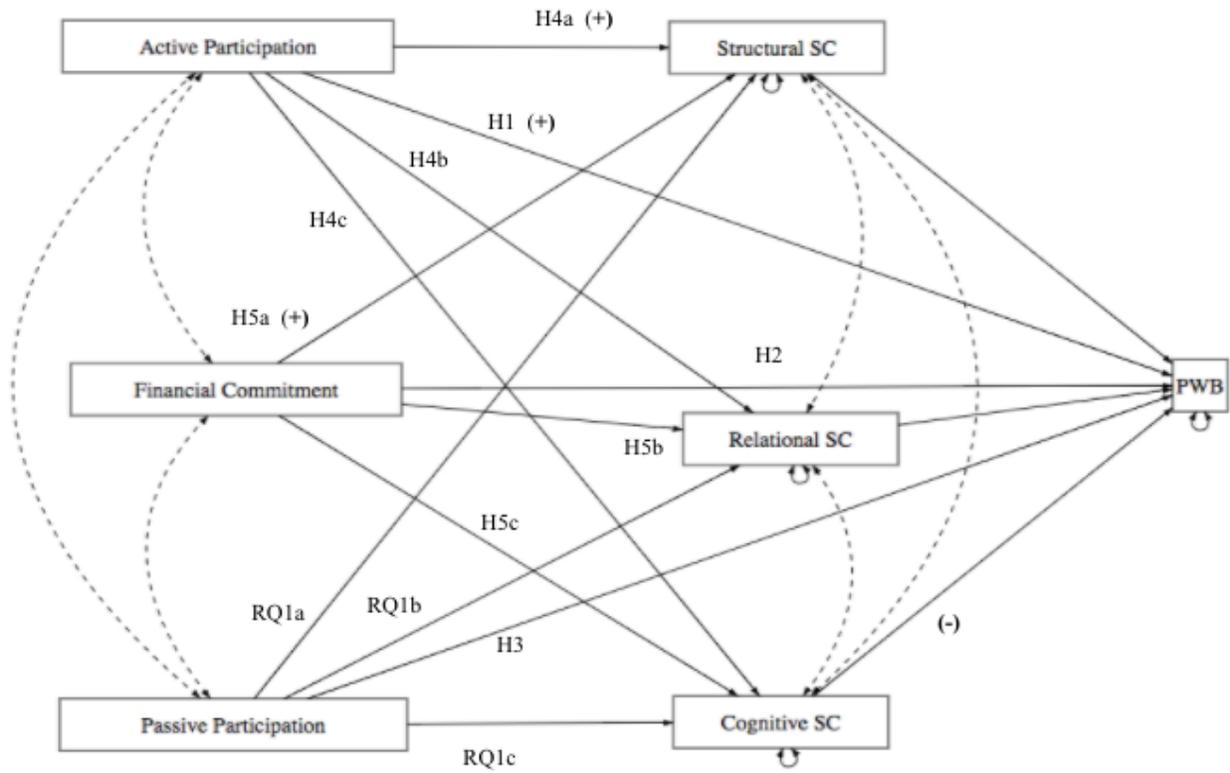

*Note.* (+) indicates positive significance while (–) indicates negative significance.

Figure 2. SEM with PSR as Mediator

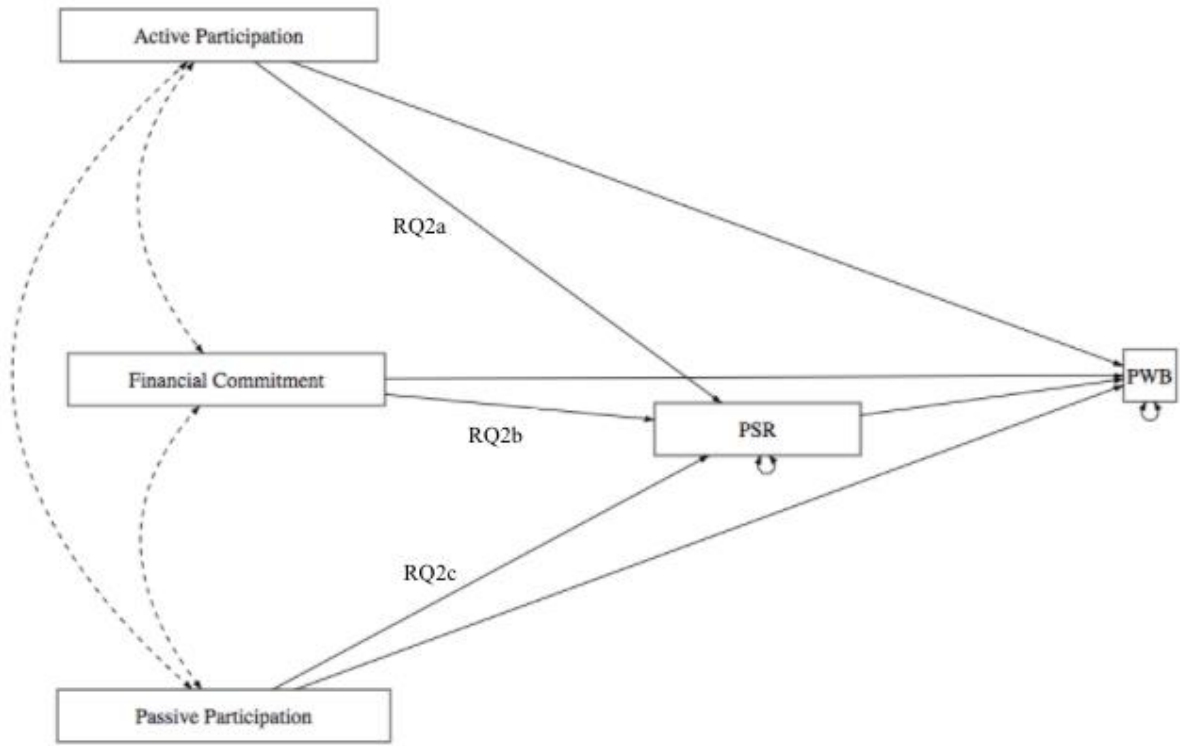

Table 1. Means, Standard Deviations, and Correlations

| Variables | M | SD | 1 | 2 | 3 | 4 | 5 | 6 | 7 | 8 | 9 | 10 | 11 |
|---|---|---|---|---|---|---|---|---|---|---|---|---|---|
| 1. Active Participation | 1.95 | 0.98 | | | | | | | | | | | |
| 2. Passive Participation | 3.59 | 1.1 | 0.52*** | | | | | | | | | | |
| 3. Financial Commitment | 21.57 | 0.45 | 0.35*** | 0.28*** | | | | | | | | | |
| 4. Structural Social Capital | 2.11 | 1.17 | 0.45*** | 0.2*** | 0.26*** | | | | | | | | |
| 5. Relational Social Capital | 3.46 | 0.93 | 0.37*** | 0.33*** | 0.28*** | 0.44*** | | | | | | | |
| 6. Cognitive Social Capital | 3.69 | 0.85 | 0.22*** | 0.22*** | 0.2*** | 0.17*** | 0.56*** | | | | | | |
| 7. Parasocial Relationship | 3.56 | 0.69 | 0.35*** | 0.27*** | 0.23*** | 0.36*** | 0.68*** | 0.52*** | | | | | |
| 8. Psychological Well-being | 3.35 | 0.85 | 0.1* | -0.01 | 0.07 | 0.2*** | 0.12* | -0.03 | 0.08 | | | | |
| 9. Time Spent | 9.59 | 11.77 | 0.23*** | 0.27*** | 0.28*** | 0.17*** | 0.23*** | 0.26*** | 0.15** | -0.12* | | | |
| 10. Perceived Offline Support | 39.96 | 10.09 | -0.03 | -0.07 | 0.09' | 0.1* | 0.11* | 0.04 | 0.06 | 0.58*** | -0.12* | | |
| 11. Health | 2.86 | 1.1 | -0.05 | -0.04 | -0.02 | 0 | 0.05 | 0.03 | -0.04 | 0.41*** | 0.02 | 0.18*** | |
| 12. Account Age | 4.16 | 2.39 | 0.08 | 0.1* | 0.19*** | 0 | 0.05 | 0.08 | -0.01 | -0.04 | 0.24*** | -0.03 | 0.05 |

*Note.* M and SD are used to represent mean and standard deviation respectively. *p<.05. **p<.01. ***p<.001.

Table 2. SEM Results with Social Capital as Mediator (N= 396)

| | Structural Social Capital | Relational Social Capital | Cognitive Social Capital | Psychological Well-being |
|---|---|---|---|---|
| | b | b | b | b |
| **Direct** | | | | |
| Active Participation | 0.528 [.382, .660] | 0.236 [.133, .329] | 0.097 [-.008, .193] | 0.089 [.010, .173] |
| Financial Commitment | 0.272 [.011, .595] | 0.262 [.080, .446] | 0.151 [-.019, .328] | 0 [-.194, .152] |
| Passive Participation | -0.073 [-.185, .036] | 0.108 [.006, .207] | 0.066 [-.025, .163] | -0.010 [-.080, .055] |
| Structural Social Capital | | | | 0.086 [.018, .154] |
| Relational Social Capital | | | | 0.055 [-.034, .147] |
| Cognitive Social Capital | | | | -.111 [-.210, -.018] |
| **Indirect** | | | | |
| Active x Structural | | | | .045 [.011, .089] |
| Active x Relational | | | | .013 [-.007, .041] |
| Active x Cognitive | | | | -.011 [-.032, 0] |
| Financial x Structural | | | | 0.023 [.002, .069] |
| Financial x Relational | | | | .014 [-.007, .053] |
| Financial x Cognitive | | | | -.017 [-.053, 0] |
| Passive x Structural | | | | -0.006 [-.023, .002] |
| Passive x Relational | | | | 0.006 [-.002, .023] |
| Passive x Cognitive | | | | -.007 [-.028, .001] |
| **Covariates** | | | | |
| Time Spent | 0.007 [-.002, .016] | 0.009 [.002, .016] | 0.015 [.008, .021] | -0.006 [-.012, -.001] |
| Perceived Offline Social Support | 0.014 [.004, .026] | 0.013 [.005, .021] | 0.006 [-.003, .014] | 0.042 [.034, .048] |
| Education | -0.071 [-.148, .014] | -0.109 [-.171, -.047] | -0.011 [-.077, .055] | 0.061 [.012, .110] |
| Health | 0.011 [-.089, .107] | 0.067 [-.013, .146] | 0.038 [-.044, .120] | 0.253 [.189, .312] |
| Sex | 0.070 [-.091, .237] | 0.178 [.047, .314] | 0.121 [.-.030, .254] | 0.019 [-.086, .128] |
| R Squared Estimate | 0.234 | 0.243 | 0.121 | 0.490 |

*Note.* Values in the square brackets indicate the 95% confidence interval for each correlation.

---

[i] Additional analyses added employment status as a covariate. 32% of respondents were employed full-time, 30% unemployed (and job seeking), 17% employed part-time, 12% were not involved in paid work (e.g. homemaker, retired, or disabled), and 9% reported "Other." Employment was not significant and only explained 1% more variance and therefore was excluded from the final analyses.